\documentclass[12pt]{iopart}
\usepackage{iopams}  
\usepackage{graphicx}  
\usepackage{amsfonts}  
\usepackage{mathrsfs}  

\newcommand{\ii}{\;\! \mbox{i} \;\!}
\newcommand{\iis}{\;\! \mbox{\scriptsize i} \;\!}

\newcommand{\bbR}{\mathbb{R}}

\newcommand{\bbN}{\mathbb{N}}
\newcommand{\Rop}{{\hat R}}
\begin{document} 
 
\title[Tunnelling rates for the nonlinear Wannier-Stark problem]
{Tunnelling rates for the nonlinear Wannier-Stark problem}
 
\author{Sandro Wimberger\dag\footnote[3]
{Corresponding author's e-mail: saw@df.unipi.it},  
Peter Schlagheck\ddag\, and Riccardo Mannella\dag} 
 
\address{\dag Dipartimento di Fisica Enrico Fermi and CNR-INFM,  
Universit\`{a} degli Studi di Pisa, Largo Pontecorvo 3, I-56127 Pisa} 
 
\address{\ddag Institut f{\"u}r Theoretische Physik,
Universit{\"a}t Regensburg, D-93040 Regensburg}
 
\begin{abstract} 

We present a method to numerically compute accurate tunnelling rates
for a Bose-Einstein condensate which is described by the nonlinear
Gross-Pitaevskii equation. Our method is based on a sophisticated 
real-time integration of the complex-scaled Gross-Pitaevskii 
equation, and it is capable of finding
the stationary eigenvalues for the Wannier-Stark problem. 
We show that even weak nonlinearities have significant effects 
in the vicinity of very sensitive resonant tunnelling peaks, which occur
in the rates as a function of the Stark field amplitude. The mean-field
interaction induces a broadening and a shift of the peaks, and the latter
is explained by analytic perturbation theory.   

\end{abstract} 
 
\pacs{03.75.-b,03.65.Xp,05.60.Gg,02.70.Bf} 

%\maketitle 
 
\section{Introduction} 
\label{eins} 

Quantum dynamics often is intriguing and counter-intuitive. A prominent
example thereof is the localisation of a wave packet
in a spatially periodic lattice induced by an additional static force:
the force can turn an extended Bloch wave (which is a solution of the
Schr\"odinger equation with a periodic potential \cite{Kittel})
to a wave packet which oscillates 
periodically in (momentum) space \cite{Kittel}.
While conceptionally simple, this well-know Wannier-Stark problem is
complicated from the mathematical point of view because the system 
is open, i.e., unbounded, and any initially prepared state
will, in the course of time evolution,
decay via tunnelling out of the periodic potential wells
\cite{nenciu,GS1997}.

Starting from the Bloch bands of the unperturbed problem (i.e., without
the static field $F=0$), the decay can be attributed to tunnelling from the
ground state band to the first excited energy band. The celebrated
Landau-Zener
theory predicts an exponential decay rate (see, for instance, 
\cite{holt,kolo} for introductory reviews):
\begin{equation}
\Gamma (F) \propto F e^{-\frac bF}\,,
\label{eq:lz}
\end{equation}
where $b$ is proportional to the square of the energy gap between the
two lowest energy bands. For experiments with cold atoms, i.e., the scenario
on which we focus in this paper, the wave packet decays very quickly
by successive tunnelling events, once it has tunnelled across the first 
band gap. This is due to the much smaller gaps of the higher energy bands
in a sinusoidal potential \cite{KK2004,BEC_pisa}.
The Landau-Zener formula (\ref{eq:lz}) cannot account for the interaction
of the Wannier-Stark levels at adjacent potential wells. Between such
adjacent lattice sites nearly degenerate Wannier-Stark levels repel
each other, which leads to a strong enhancement of the tunnelling decay.
These resonant tunnelling events result in pronounced peaks in the 
rates as a function of the inverse field amplitude $1/F$, on top
of the of the global exponential decay described by (\ref{eq:lz})
\cite{kolo,koloPRL}.

\begin{figure} 
\centering 
\includegraphics[height=10cm]{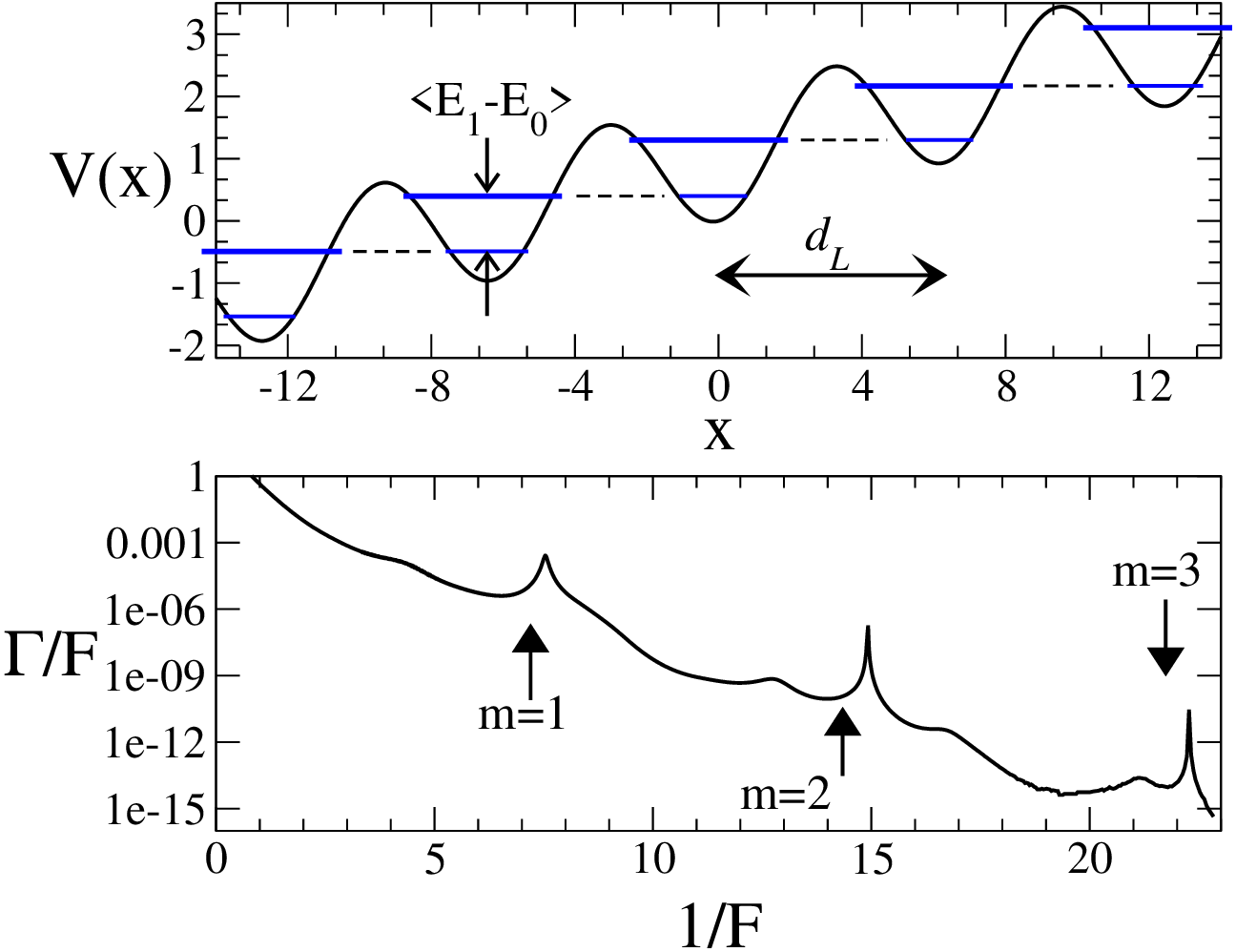} 
\caption{(a) Schematic sketch of nearly degenerate Wannier-Stark
             levels (thin line: ground state levels; thick line:
             first excited levels in each well) in a potential of the form
             $V(x)=V_0 \sin^2(x/2) + Fx$.
         (b) tunnelling rates $\Gamma$ for $V_0=2$ as a function of
             the inverse Stark field amplitude $1/F$. 
             The resonant tunnelling leads
             to the pronounced peaks, which lie approximately at 
             $F \approx \langle E_{1}-E_{0}\rangle/(2\pi m)$ 
             (with integer $m$). These estimates
             (marked by the arrows)
             are slightly modified by field-induced level shifts.
}  
\label{fig:1}. 
\end{figure}

Figure \ref{fig:1}(a) shows two Wannier-Stark levels on each lattice
site. The levels within either of the two ladders are separated 
by $mFd_L$ in energy, where $d_L\equiv 2\pi$ denotes the lattice period and
the integer $m$ counts the number of 
sites in-between two energy levels of the same ladder 
\cite{Kittel,nenciu}.
The decay rates for non-interacting particles in the periodic potential
$V(x)=V_0 \sin^2(x/2) + Fx$ can be computed from the Wannier-Stark 
spectrum, e.g.,
by using the numerical method described in \cite{kolo}. Figure \ref{fig:1}(b)
presents the rate $\Gamma$ as a function of $1/F$. The maxima occur
when $mFd_L$ is close to the difference in energy 
$\langle E_{1}-E_{0}\rangle$ between the first two energy
bands (averaged over the fundamental Brillouin zone in momentum space)
of the unperturbed ($F=0$) problem \cite{kolo,glutsch2004}.
The actual peak positions are slightly shifted with respect to this
simplified estimate [marked by arrows in Fig. \ref{fig:1}(b)], 
owing to a field-induced level shift close to the avoided crossings
of the levels \cite{koloPRL}.

Exceptional experimental control is possible nowadays with 
Bose-Einstein condensates (BEC) whose initial conditions in coordinate
and momentum space can be adjusted with unprecedented precision. 
With the help of a BEC, sensitive tunnelling phenomena were
studied in time-dependent systems \cite{BEC_dynamic}, as well as in static
potentials \cite{BEC_mott}. Here we are interested in tunnelling in 
the Wannier-Stark problem where the impact of the intrinsic atom-atom 
interactions in the BEC has been studied in several recent experiments 
\cite{BEC_pisa,BEC_bloch,BEC_asym,BEC_flo}. In those experiments,
the difficulty in understanding quantum transport
processes, such as coherent tunnelling, originates from
the complex interplay between classical transport in the underlying 
phase space, quantum interference effects, and the many-particle interactions.

In a typical experiment with a BEC, where the number of atoms in the
condensate is large and the atom-atom interaction is rather small,
the Gross-Pitaevskii equation (GPE) describes the condensate in very
good approximation \cite{BEC_books}. 
Recently, some of us proposed a concrete experimental
scenario to measure the impact of a mean-field interaction potential
(which in the GPE takes into account of the atomic collisions) on the
tunnelling in the Wannier-Stark problem \cite{pisa_WS}. More specifically, the 
interaction-induced modification of the resonant tunnelling peaks was
studied, and it was found that the peaks [such as the ones in Fig. 
\ref{fig:1}(b)] are washed out for a large enough -- but still
experimentally feasible -- interaction strength.
As discussed in \cite{pisa_WS}, for a finite mean-field nonlinearity,
the concept of decay rates is not as well defined as in the case of 
non-interacting particles. The reason is that the weight of the nonlinear
term, which is proportional to the condensate density, varies in time.
Hence,
the probability that an initially prepared state will stay in
the preparation region does not follow a simple exponential law. In other
words, the nonlinear interaction decreases as the condensate escapes via
tunnelling and, as a consequence, the decay rates can be defined
only locally in time.

In this paper we want to discuss how the problem of 
defining proper decay rates can be solved.
One way to define a time-independent, global tunnelling rate is to
renormalise the density of the condensate in the preparation region
(e.g., in the central potential well of the periodic lattice) continuously,
such that the average density remains constant in time. For experimentally
realisable nonlinearities \cite{BEC_pisa,BEC_bloch,BEC_asym,chu}, 
this approach results in a mono-exponential
decay of the survival probability, and in consequence in
a reasonable definition of the decay rate. The corresponding resonance
states are characterised by their stationary asymptotics, much in the
same way as the stationary solutions of a linear scattering problem (i.e,
described by the linear Schr\"odinger equation) \cite{kolo}.
We solve numerically the well-posed problem of finding the resonance states,
using the method of complex scaling. The theoretical background to treat the
nonlinear interaction in the GPE in a consistent way was recently laid in
\cite{SP2004}. We use a modified version of this method, 
with crucial extensions on the
algorithmic side, which proved necessary to stabilise the computations
for more complicated potentials than the single-well potential
exemplarily treated in \cite{SP2004}.

We review the defining equations of the nonlinear Wannier-Stark problem and the
complex scaling technique of \cite{SP2004} in the following section 
\ref{zwei}, where we also describe our numerical algorithm in detail, c.f.,
subsection \ref{algo}. Section \ref{drei} presents our central results on the
decay rates in the vicinity of the resonant tunnelling peaks for experimentally
relevant nonlinearities. Section \ref{concl} finally concludes the paper.

\section{The nonlinear Wannier-Stark problem}
\label{zwei}

We use the one-dimensional GPE to model the temporal evolution
of a BEC loaded into a spatially periodic optical lattice 
potential and subjected to an additional static force $F$:
\begin{equation}
\ii \frac{\partial}{\partial t}\psi (x,t) =
\left[-\frac12 \frac{\partial^2}{\partial x^2} 
+ V_0 \sin^2 \left( \frac{x}{2} \right) + F x + g \left| \psi(x,t) \right|^2
\right] \psi(x,t) \;.
\label{eq:GP}
\end{equation}
$\psi(x,t)$ represents the condensate
wave function, and we used the dimensionless quantities
$V_0=V_{\rm SI}/E_{\rm B}, F=F_{\rm SI}d_L/(2\pi E_{\rm B}),
g=g_{\rm SI}d_LN/(2\pi E_{\rm B})$. The characteristic length scale is the
lattice period $d_L$, i.e., $x=x_{\rm SI}2\pi/d_L$, the Bloch energy
is $E_{B}=(\pi \hbar/d_L)^2/M$, with atomic mass $M$, the number of
atoms $N$, and the (from three to one spatial dimensions) 
rescaled nonlinearity parameter $g_{\rm SI}$ (see \cite{O1998} 
for a definition of $g_{\rm SI}$, where also the regime
of validity of the one-dimensional approximation is discussed in detail).

Since $V(x)=V_0 \sin^2(x/2) +Fx \to -\infty$ for $x \to -\infty$,
any state initially prepared in the optical lattice will
escape via tunnelling. We search for the resonance state $\psi_g$
which solves the stationary version of Eq. (\ref{eq:GP})
\begin{equation}
H[\psi_g] \psi_g = E_g \psi_g \;,
\label{eq:eig}
\end{equation}
for the eigenvalue $E_g=\mu_g - i \Gamma_g/2$, and the Hamiltonian
\begin{equation}
H[\psi]= - \frac12 \frac{\partial^2}{\partial x^2} + V(x) + g |\psi (x)|^2
\end{equation}

To render the problem posed by Eq. (\ref{eq:eig}) meaningful we demand
that the condensate wave function remains normalised around the 
initially prepared state, i.e., around $x\approx 0$:
\begin{equation}
\int_{-x_n}^{x_n} \; dx \left| \psi_g(x) \right|^2 = 1 \;.
\label{eq:renorm}
\end{equation}
The boundaries $x_n$ must be chosen in a reasonable way, and we chose
$x_n=\pi$ (so the probability to stay in the central well around $x=0$ remains
one \cite{KMPS2003}). 
We verified that slightly different choices of the boundary $ \pi \lesssim x_n 
< 3\pi/2$ led to eigenvalues which did not change on the 
significant digits given below in section \ref{drei}. 

We discuss now the renormalisation condition (\ref{eq:renorm})
and its consequences. In practice such a condition may be realised
by the presence of a source term which constantly supplies a condensate
flow \cite{PRS2005}. Experimentally such a scenario could be achieved
by constantly reloading the central well with coherent BEC matter. 
Transport experiments of such kind could be realised with the help
of optical tweezers \cite{tweez}, atomic conveyer belts \cite{conv}, or
microscopic guides for ultracold atoms \cite{schmied2003}. 
Any realisation may introduce additional modifications in the 
temporal evolution of the  decaying system, which go beyond our 
simplified assumption of renormalisation.
Such modifications, e.g., the relaxation of added particles in the periodic
lattice potential, depend on the specific realisation. We expect, however,
that the asymptotic decay will be hardly affected by such processes, the
time scales of which should be relatively short and of the order of the 
period of oscillations in the potential wells.

On the other hand, if the condensate wave function tunnels out of the
central well without sudden changes of its shape, the
time-dependent atomic population $N(t)$ 
inside the well decays according to the relation \cite{SP2004,CHM2005}
\begin{equation}
\frac{dN(t)}{dt} = - \Gamma_{g(t)} N(t) \;.
\label{eq:ndecay}
\end{equation}
Assuming that the decay rate adiabatically adjusts 
itself to the time-dependent value
$\Gamma_{g(t)}$, with $g(t) \propto N(t)$,
Eq. (\ref{eq:ndecay}) can be solved for a given initial number of atoms
$N(0)$ in the condensate. Knowing the ``local''
rates $\Gamma_{g(t)}$ for $0 \leq |g| \leq |g(0)|$ allows us then to 
compute the actual survival probability in the central well, which 
in \cite{pisa_WS} was obtained differently by a brute force integration of the
time-dependent GPE (\ref{eq:GP}). 

We emphasise that the setup studied in Ref. \cite{pisa_WS} bears some
crucial differences to the problem posed here, which is based on condition 
(\ref{eq:renorm}). In \cite{pisa_WS} the short time behaviour of the
relaxed ground state (for $F=0$ in the periodic potential and in the 
presence of additional harmonic confinements) was predicted for the 
three dimensional Wannier-Stark problem. The approach presented here is
capable to determine, via Eq.~(\ref{eq:ndecay}), the decay only for
\emph{single} resonance states according to the above arguments.
Although such resonance states are typically distributed over many lattice
sites, they do not provide a prediction for the decay of a general initial
state (which could be composed of contributions from many adjacent wells),
simply because the superposition principle does not apply for the nonlinear
GPE (\ref{eq:GP}).

In this paper we want to compute {\it directly} the precise 
decay rates $\Gamma_{g}$ of a single resonance state
using the complex scaling method, which is described in the 
following subsection.

\subsection{Complex scaling} 
\label{scaling} 

For the linear problem with $g=0$, one of the standard techniques
to compute resonance states numerically is the complex scaling
method (which goes back to \cite{BC1971}, and is reviewed, for instance, 
in \cite{KKH1989}).
Applying the renormalisation condition (\ref{eq:renorm}) allows us
to use this method to find the stationary eigenstates and the 
corresponding eigenvalues, see Eq. (\ref{eq:eig}). Without this
condition, the nonlinear interaction term would vary in time, and a 
stationary state would not exist because of the tunnelling decay.

An additional problem when dealing with the nonlinear term in the
GPE arises from the method of complex scaling itself. The problem
of defining the complex conjugate of the wave function $\psi(x)$
is described in \cite{SP2004,MC2005}, and was solved in \cite{SP2004}.
Usually, the scaling transformation is defined as follows
\begin{equation}
\psi(x) \to \psi^{\theta}(x) \equiv \Rop(\theta) \psi(x) \equiv
e^{\iis \theta/2} \psi(xe^{\iis \theta}) \;,
\label{eq:scal1}
\end{equation}
where the pre-factor is just a phase depending on the dimensionality
of the problem (here we treat only the one-dimensional case). $\theta$
is a real rotation angle, and the eigenvalues should not depend on
it \cite{BC1971,KKH1989}, which is a useful fact for testing convergence.
To evaluate the nonlinear term $|\psi|^2=\psi^*\psi$ away from the
real coordinate (or $x$) axis, we need to define a generalised complex
conjugate $\overline{\psi}$ which reduces to 
$\overline{\psi}(x) = \psi (x)^*$ for $x \in \bbR$.
Applying the complex scaling transformation to $\overline{\psi}$
\begin{equation}
\overline{\psi(x)} \to \overline{\psi}^{\theta}(x) 
\equiv \Rop(\theta) \overline{\psi}(x) \equiv
e^{\iis \theta/2} \overline{\psi}(xe^{\iis \theta}) \;,
\label{eq:scal2}
\end{equation}
we see that $\overline{\psi}^{\theta}$ can be obtained from $\psi^{\theta}$
via the relation:
\begin{equation}
\overline{\psi}^{\theta}(x) =
\Rop(\theta) \left( \Rop(-\theta) \psi^{\theta} \right)^* (x)\;.
\label{eq:scal-bar}
\end{equation}

The analytic continuation of Eq. (\ref{eq:eig}) to the complex domain
can now be stated as
\begin{equation}
H^{\theta}[\psi_g^{\theta}] \psi_g^{\theta} = E_g \psi_g^{\theta} \;,
\label{eq:eig-com}
\end{equation}
with 
\begin{equation}
H^{\theta}[\psi_g^{\theta}] = -\frac12 \frac{\partial ^2}{\partial x^2}
e^{-\iis2\theta} + V(xe^{\iis \theta}) + g_{\theta} 
\overline{\psi}_g^{\theta}(x) \psi_g^{\theta} (x) \;.
\label{eq:ham-com}
\end{equation}
The nonlinear interaction strength is defined here as 
$g_{\theta} = ge^{-\iis \theta}$ to compensate for 
the two identical phase factors
$e^{\iis \theta/2}$ of $\psi^{\theta}$ and $\overline{\psi}^{\theta}$.

\subsection{Numerical solution and propagation algorithm}
\label{algo} 

In the linear case with $g=0$, the complex eigenvalue problem
of the form Eq.~(\ref{eq:eig-com}) 
is usually solved by representing the complex-scaled
Hamiltonian in a suitable basis and final matrix diagonalisation 
\cite{complexscal}.
For $g\neq 0$, the corresponding problem to find the eigenvalues can
be solved only by implicit methods, since $H^{\theta}[\psi^{\theta}]$
explicitly depends on the wave function.

We solved Eq. (\ref{eq:eig-com}) by searching for the ground-state
solution in a self-consistent manner. Starting with an initial guess
for the wave function $\psi^{\theta}(x,t=0)$, we evolved in real-time 
the grid representation of $\psi^{\theta}(x,t)$, i.e.,
\begin{equation}
\psi^{\theta}(x,t) = \sum_{j=-n}^n c_j(t) \chi_j(x) \;,
\label{eq:grid}
\end{equation}
with the box functions
\begin{eqnarray}
\chi_j (x)
= \left\{
         \begin{array}{r@{\quad , \quad}l}
         1/\Delta_x
  & \left|x/\Delta_x - j \right| < 1/2
                                 \\ 
0 & \mbox{\small otherwise}\;,

\end{array} \right.
\label{eq:box}
\end{eqnarray}
and a suitable grid spacing $\Delta_x$.

The time-propagation was performed by a sequential application of two
different integration methods. First, we used a sequence of Crank-Nicholson
steps \cite{numres}, i.e.,
\begin{equation}
\left( 1 + \ii H^{\theta}\Delta t/2\right) 
\psi^{\theta}(x,t+\Delta t/2) = 
\left( 1 - \ii H^{\theta}\Delta t/2\right) 
\psi^{\theta}(x,t-\Delta t/2)\;.
\label{eq:CN}
\end{equation}
The Crank-Nicholson method has the advantage of preserving the norm
of the wave function, but the disadvantage that it treats all modes
equally. Since we are interested in the ground state, we iterated in 
a second stage the explicit relation
\begin{equation}
\psi^{\theta}(x,t+\Delta t) = 
\left( 1 - \ii H^{\theta}[\psi^{\theta}]\Delta t \right) 
\psi^{\theta}(x,t)\;.
\label{eq:expl}
\end{equation}
The latter method, which still corresponds to a real-time integration of the
complex scaled Gross-Pitaevskii equation, tends to suppress the higher modes
\cite{numres} and leads to a faster stabilisation of the numerical solution of
Eq.~(\ref{eq:eig-com}) in comparison with the Cranck-Nicholson method 
(\ref{eq:CN}).
For {\em each} time step $t \mapsto t + \Delta t$,
we self-consistently solved
Eq.~(\ref{eq:expl}) by using the left side of Eq.~(\ref{eq:expl}) 
to approximate the nonlinear term 
$g_\theta\overline{\psi}^{\theta}\psi^{\theta}$. Three to five such
self-consistent iterations proved sufficient for a
stable and reliable time propagation. The second derivative appearing
in $H^{\theta}$ was approximated by a finite difference representation 
(in other words we applied the ``forward time centred space'' 
representation \cite{numres} to solve the GPE). This leads to a 
tridiagonal Hamiltonian matrix, which significantly simplifies the
implementation of both propagators (\ref{eq:CN}) and (\ref{eq:expl}).

For evaluating  $\overline{\psi}^{\theta}(x,t)$, we used the method
described in detail in \cite{SP2004}, which produced reliable numerical
results also for our Wannier-Stark problem. Briefly speaking, we
represent $\psi^{\theta}(x,t)$ in a basis set of Gaussians with increasing
variance for increasing $|x|$.
The Gaussian basis is thus well-behaved at the boundaries of our grid, which
allows us a numerically stable back-rotation to the real domain in $x$.
At the end, $\overline{\psi}^{\theta}(x,t)$ is re-expressed again in the
grid basis. The necessary matrix-vector multiplications are fast since the
number of vectors in the Gaussian set can typically be chosen much smaller than
the number of grid points in the spatial domain. Furthermore, the
transformation matrices are effectively banded, which reduces the numerical
effort (note that we computed $\overline{\psi}^{\theta}(x,t)$ from
$\psi^{\theta}(x,t)$ for {\em each} time step 
$t \mapsto t + \Delta t$ to ensure stable convergence).

\section{Results and discussion}
\label{drei} 

In the following, we present our results on the tunnelling rates of resonance
states [c.f. Eq. (\ref{eq:eig-com})] in the Wannier-Stark problem as sketched 
in Fig. \ref{fig:1}(a). 
Without loss of generality we kept fixed the potential depth 
$V_0=2$ in Eq.~(\ref{eq:GP}), which corresponds to an optical lattice
with a maximal amplitude of $16$ photon recoil energies \cite{KK2004,BEC_pisa}.
We were particularly interested in studying the impact of the nonlinear
term in Eq.~(\ref{eq:GP}) on the resonant tunnelling peaks of Fig. 
\ref{fig:1}(b). Using the method described in the previous section
we chose $\theta=0.01\ldots 0.02$ (where we found stable eigenvalues
which are not dependent on $\theta$ in this range), and a grid spacing
$\Delta_x=0.02\ldots 0.05$ for $-100 \leq x \leq 100$. The integration
time step was $\Delta t = 2.5\times 10^{-3}$ for $|g| < 0.2$ and reduced
to $\Delta t = 2\times 10^{-3}$ for larger $|g| \geq 0.2$ and
the region of small $F<0.2$, while the
maximal integration time for finding one eigenvalue was $t_{\rm max}=300$. 

As expected, it is very difficult to find the correct eigenvalue
close to a resonant tunnelling peak because of two reasons:
(i) the rates $\Gamma$ vary dramatically around the peak due to the
    close degeneracy of the Wannier-Stark levels in adjacent wells, and
(ii) the rates are rather small $10^{-13} < \Gamma \lesssim 2\times 10^{-3}$ 
    at small fields $0.05 < F < 0.2$.

Therefore, we focused on the first (i.e., with $m=1$) peak in Fig.
\ref{fig:1}(b), where the rates remain $\Gamma \gtrsim 10^{-5}$.
Moreover, we improved the stability of the integration by starting with
parameters in a stable regime (where we easily found stable, fast
converging eigenvalues) and adiabatically changing the two parameters
$F$ and $g$ into less stable parameter regimes. In our case, for $V_0=2$,
the stable regime is above $F>0.25$ for not too large nonlinearities
$|g| \lesssim 0.5$ (with optimal stability properties for $g=0$). 

\begin{figure} 
\centering 
\includegraphics[height=7.5cm]{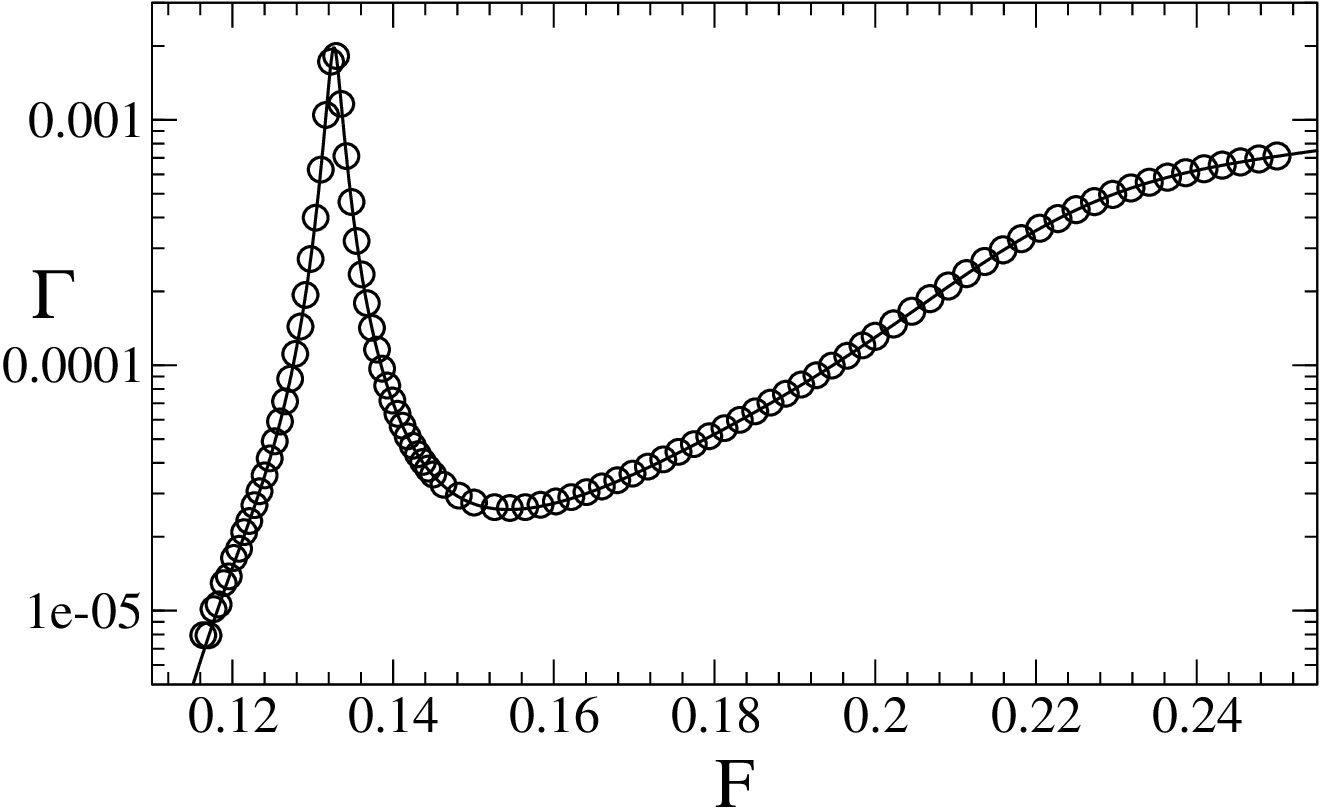} 
\caption{ 
Comparison between the tunnelling rates around the first resonant
tunnelling peak from Fig. \ref{fig:1}(b), obtained for $g=0$ by direct
diagonalisation of the problem (solid line) and by our complex
scaling algorithm (circles). 
}  
\label{fig:2}. 
\end{figure}

We tested our results in three different and independent ways.
First, we compared them for $g=0$ with the spectra of the linear 
Wannier-Stark problem, which can be computed by a standard diagonalisation
of $H^{\theta}_{g=0}$ \cite{kolo,complexscal}. 
Figure \ref{fig:2} shows the good agreement between the 
data set obtained by our integration of the complex
scaled, linear version of Eq. (\ref{eq:GP}) and the data presented already
in Fig. \ref{fig:1}(b).

Secondly, for moderate nonlinearities $|g| \leq 0.5$, we computed
for the {\it un-scaled} problem (\ref{eq:GP}) the survival probability
\begin{equation}
P_{\rm sur} (t)\equiv \int_{-p_c}^\infty \; dp |\hat{\psi}(p,t)|^2 
\approx N(t) \;, 
\label{eq:sur}
\end{equation}
with $p_c \gtrsim 3$ photon recoils (such as to cover the support
in momentum space of the initial state prepared in the spatially 
periodic lattice potential). $P_{\rm sur} (t)$ was introduced
for the Wannier-Stark problem in \cite{pisa_WS} and characterises the 
out-coupled loss in momentum space, which corresponds to the part
of the condensate which has tunnelled through the potential.
We integrated Eq.~(\ref{eq:GP}) constantly applying the renormalisation
condition (\ref{eq:renorm}), and computed the survival probability
(\ref{eq:sur}) with the wave function
\begin{equation*}
\hat{\psi}(p,t) \approx  \frac{\hat{\psi}_{\rm renorm}(p,t)}{\prod_{j=1}^{K}
\int_{-\pi}^{\pi} \; dx \left| \psi(x,jt/K) \right|^2}\;,
\end{equation*}
for the discretised times $jt/K$ ($j=1,2,\ldots,K$) and large $K \in \bbN$.
Here $\psi(x,jt/K)$ denotes the propagated wavefunction immediately 
before applying
the renormalisation ($\psi(x,jt/K)$ is renormalised afterwards and 
propagated up to time $(j+1)t/K$). $\hat{\psi}_{\rm renorm}(p,t)$
represents the Fourier transform of the renormalised wavefunction at the end
of the complete propagation.
The decay rates $\Gamma_{\rm fit}$ were obtained by a direct 
mono-exponential fit to the temporal decay of $P_{\rm sur} (t)$. Table 
\ref{table:1} highlights the good agreement with the rates computed
by the complex-scaling method.

\begin{table} \begin{center} \begin{tabular}{|l||l|l|l|} 
\hline
 $g$  & F   & $\Gamma_{\rm fit}$ & $\Gamma_{\rm CS}$ 
\nonumber \\ \hline\hline
0     & 0.5  &  $ 1.94 \pm 0.01 \times 10^{-2}$ &  $ 1.941 \times 10^{-2}$ 
\nonumber \\ \hline
0.1   & 0.5  & $2.18 \pm 0.01 \times 10^{-2}$ & $2.180 \times 10^{-2}$ 
\nonumber \\ \hline
0     & 0.25 &  $ 7.2 \pm 0.1 \times 10^{-4}$ &  $ 7.2 \times 10^{-4}$ 
\nonumber \\ \hline
0.1   & 0.25 & $8.4 \pm 0.1 \times 10^{-4} $ &  $ 8.4 \times 10^{-4}$ 
\nonumber \\ \hline
0.2   & 0.25 & $9.7  \pm 0.1 \times 10^{-4}$ &   $9.7 \times 10^{-4}$ 
\nonumber \\ \hline
0.25  & 0.25 & $1.04 \pm 0.02 \times 10^{-3}$ & $1.04\times 10^{-3}$ 
\nonumber \\ \hline
0.5   & 0.25 & $1.45 \pm 0.03 \times 10^{-3}$ & $1.48 \times 10^{-3}$  
\nonumber \\ \hline
0.2   & 0.15 & $3.0  \pm 0.2  \times 10^{-5}$ & $ 2.9 \times 10^{-5}$ 
\nonumber \\ \hline
0.2   & 0.13125 & $ 5.7 \pm 0.3 \times 10^{-5}$ & $ 5.7 \times 10^{-5}$ 
\\ \hline
\end{tabular} 
\caption{Comparison between the tunnelling rates for $V_0=2$ obtained by
the complex scaling method ($\Gamma_{\rm CS}$) and by the integration
of the GPE ($\Gamma_{\rm fit}$; integration time up to 100 Bloch 
periods; the integration was performed on a large grid that 
covered the full extension of the tunnelled and subsequently 
accelerated part of the wave function, without the use of any cutoff or 
absorbing boundary conditions). 
Because of the restriction in the integration time, $\Gamma_{\rm fit}$ 
carries the shown error, whilst the complex scaling method allows us
to compute the rates $\Gamma_{\rm CS}$ with an absolute accuracy of 
at least $10^{-6}$ for 
$F \geq 0.15$, and $10^{-5}$ for $F$ down to $\gtrsim 0.12$. 
}
\label{table:1}
\end{center} \end{table}

As a final test of our results, we constantly monitored the
quality of the computed eigenvalues by evaluating
$|(H^{\theta}[\psi^{\theta}]-E) \psi^{\theta}|$, which in all
cases had to be $\lesssim 10^{-8}$ for not rejecting the eigenvalue. This
boundary was chosen such as to be more than two orders of magnitude 
smaller than the smallest tunnelling rates which we computed.

\begin{figure} 
\centering 
\includegraphics[height=10cm]{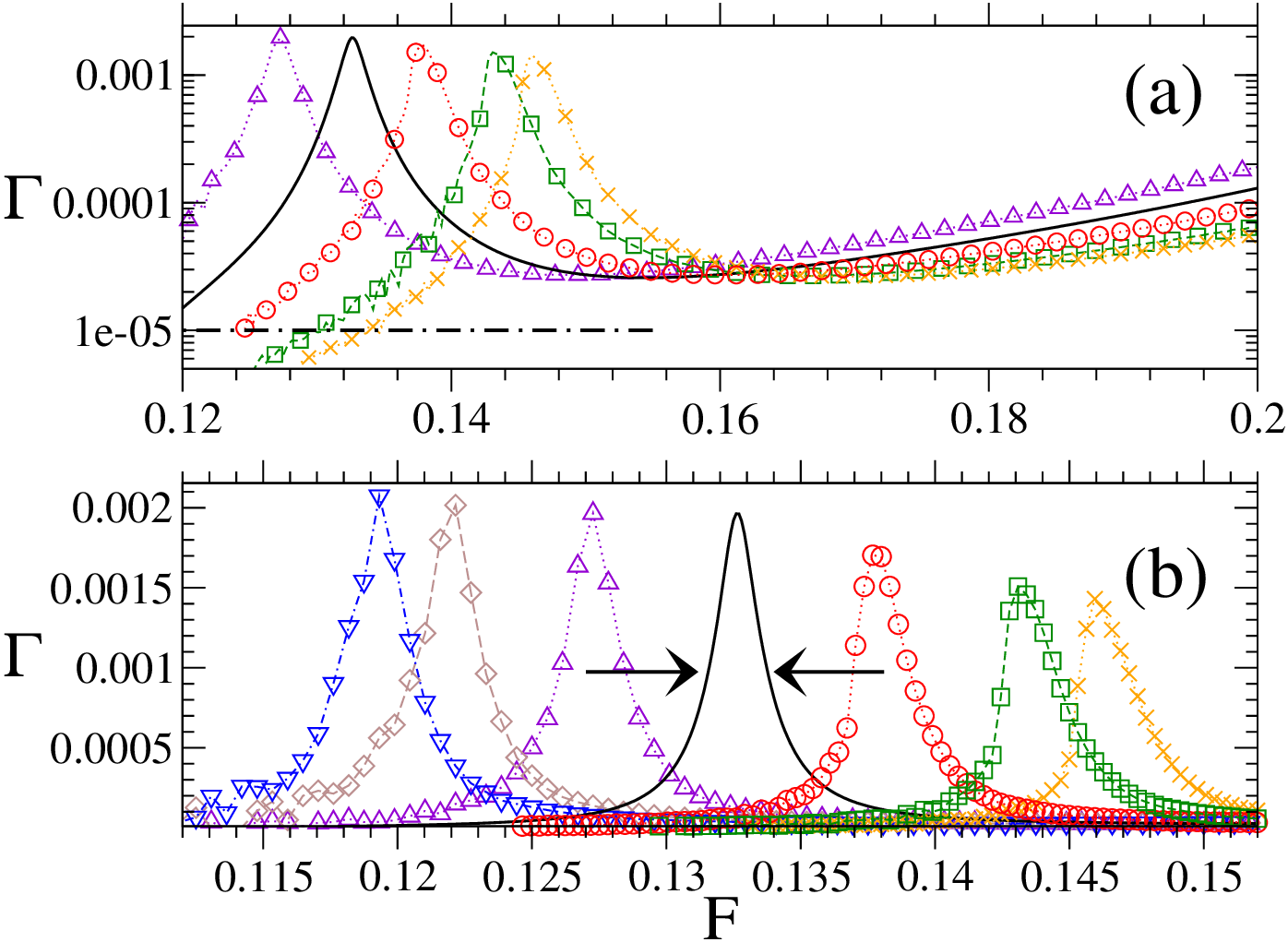} 
\caption{
Decay rates $\Gamma$ obtained for $V_0=2$ and the range
of nonlinearities $-0.25 \leq g \leq 0.25$ as a function of the
Stark field amplitude $F$ [(a) logarithmic and (b) linear scale
on y-axis]: $g=-0.25$ (crosses), $g=-0.2$ (squares), $-0.1$ (circles),
$0$ (solid line), $0.1$ (pyramids), $0.2$ (diamonds; only in (b)), and
$0.25$ (inverse pyramids; only in (b)). The nonlinearity systematically shifts
the peak centres and slightly broadens the peaks. Their width 
we defined as the full peak width at half maximum, which
is marked by the arrows in (b) for the $g=0$ peak.
The dot-dashed line in (a) shows the lower bound for converged
$\Gamma \gtrsim 10^{-5}$ on the left side of the resonant
tunnelling peaks, i.e., for very small $F\lesssim 0.12$, 
where convergence is very hard to achieve even at small $g \simeq 0.1$
(in this parameter range, i.e., at the left side of the 
peaks where $\Gamma$ changes abruptly by about two orders of magnitude,
the propagation according to Eqs.~(\ref{eq:CN}) and
(\ref{eq:expl}) becomes unstable).
}
\label{fig:3}
\end{figure}

Our central results are reported now in Fig. \ref{fig:3}.
There we observe two effects which are induced by the presence of the
nonlinear interaction term in Eq. (\ref{eq:GP}): (I) the resonant
tunnelling peak shifts systematically with increasing $g$ as a function
of the Stark field amplitude $F$. (II) the peak width slightly increases
as $|g|$ increases away from zero.

The slight broadening goes along with a small increase in the
height of the peaks with increasing nonlinearity $g$.
Such a destabilisation of the condensate for $g>0$, more precisely of
the decay in the survival probability 
$P_{\rm sur}(t)$, has already been observed in \cite{pisa_WS}. The
ratio of the difference in the height and the difference in the
peak width (measured at the half of the peak height, see Fig. \ref{fig:3})
is roughly constant as a function of the nonlinearity $0< g\lesssim 0.25$.
The broadening and the change in height of the peak are caused
by two different, but simultaneously acting mechanisms. 
The nonlinear mean-field term in Eq. (\ref{eq:GP}) partially lifts 
the degeneracy of the Wannier-Stark levels 
[as sketched in Fig. \ref{fig:1}(a)] by smearing them out. This
qualitatively explains the slight broadening of the peak 
with increasing $|g|$. 
In addition, the peak maximum becomes systematically larger as $g$ 
increases in Fig.~\ref{fig:3},
because the condensate is destabilised (stabilised) by an increasingly 
repulsive (attractive) nonlinearity.

The shift of the peak maximum can be estimated by first-order perturbation
theory, which predicts the following shift in energy of the levels 
with respect to the linear case with $g=0$:
\begin{equation}
\Delta E \approx g \int_{-x_c}^{x_c} \; dx 
         \left| \psi_g     \right|^2
         \left| \psi_{g=0} \right|^2 \;.
\label{eq:shift-1}
\end{equation}

For the moderate nonlinearities realised in experiments 
\cite{BEC_pisa,chu,wieman2001}, i.e., $|g|\lesssim 0.5$, 
the overlap integral is nearly independent of $g$, and the major contribution
comes from the central potential well where the condensate is 
localised initially. Hence we approximate further
\begin{equation}
\Delta E \approx g \int_{-\pi}^{\pi} \; dx 
         \left| \psi_{g=0}    \right|^4 \;,
\label{eq:shift-2}
\end{equation}
which in turn leads to a shift in the position on the $F$ axis
corresponding to $\Delta E \approx 2\pi \Delta F$. 
Taking into consideration that the probability in the central well remains
normalised [condition from Eq. (\ref{eq:renorm})], Eq. (\ref{eq:shift-2})
corresponds to the energy shift induced by the nonlinearity which was
baptised ``frequency pulling'' in \cite{TGZ2003}, because it leads to a
phase dispersion of the Bloch oscillations in the wells.
With (\ref{eq:shift-2}) we arrive at the general result that the 
following ratio is approximately constant:
\begin{equation}
\frac{2\pi \Delta F }{|g|} \approx \int_{-\pi}^{\pi} \; dx 
                         \left| \psi_{g=0} \right|^4 \approx 0.37 \;.
\label{eq:shift-3}
\end{equation}
This estimate proved to be valid with a maximal relative deviation
of less than $18\%$ with respect to the shifts observed in 
Fig. \ref{fig:3}.
Since the integral $\int_{-\pi}^{\pi} \; dx | \psi_{g} |^4$ is constant
up to the third digit for all $|g|\leq 0.25$, 
we conclude that the first-order perturbation 
correction is not enough to describe the shifts more quantitatively.

While a repulsive mean field ($g>0$) enhances the tunnelling rate
far away from the $g=0$ peak, an attractive interaction ($g<0$) stabilises 
the decay sufficiently far away from all the peaks. 
This is the case, e.g., for $F > 0.16$ in Fig.~\ref{fig:3}(a), where $\Gamma$
systematically decreases with decreasing $g$.
The symmetric displacement of the peak with respect to the sign
of $g$ reflects the symmetry of the Bloch band model, in which
tunnelling from the first excited band back to the ground band
can be interpreted as the converse process but with a sign change in
$g$ \cite{BEC_asym}. 
The same qualitative behaviour of enhancement ($g>0$) and stabilisation
($g<0$) was observed in the short-time evolution of the three dimensional
Wannier-Stark problem \cite{pisa_WS}. Apart from the conceptional difficulty
of decomposing a solution of the nonlinear GPE (\ref{eq:GP}) into
contributions from many adjacent wells (see discussion in section 
\ref{zwei}), the observed washing out of the peak structure 
in Ref.~\cite{pisa_WS}
is a direct consequence of an effectively moving peak as $|g|$ diminishes
monotonously (c.f. Fig. \ref{fig:3}) with decreasing density in the wells.

To conclude this section, we briefly compare our results to
other recent works which investigated the impact of a mean-field interaction
of the GPE type on quantum mechanical decay processes.
We emphasise, however, that such a comparison can only be qualitative,
since such works \cite{CHM2005,MC2005,MCMB2004,WMK2005} 
typically treat much simpler model
potentials than our Wannier-Stark problem with close level degeneracies.
The common feature of our work and the results presented in
\cite{CHM2005,MC2005,MCMB2004,WMK2005} is that an increasing
nonlinearity typically enhances the decay in the one-dimensional problem.
Systematical shifts in the chemical potential of resonance states (induced by
the interaction term) were analytically studied in \cite{WMK2005}
for a delta-shell potential. Such shifts correspond to our perturbative
estimate in Eqs.~(\ref{eq:shift-2}) and (\ref{eq:shift-3}).

Particularly, we compared the results obtained from our method
(see section \ref{zwei}) with the description of \cite{MC2005}
where a nonlinear equation was introduced which differs from ours 
[see Eq. (\ref{eq:ham-com})] in the treatment of the nonlinear term. 
In \cite{MC2005} the nonlinearity is of the form
$g_\theta (\psi^{\theta})^3$, and we found that such a nonlinear term
leads to {\it different} decay rates than the ones we computed
from either our complex scaling method or from the real-time 
integration of the un-scaled GPE (\ref{eq:GP}) and subsequent fits
to $P_{\rm sur}(t)$. We conclude that a treatment based on complex scaling
-- for a condensate within the standard GPE description and generally
complex-valued wave functions -- 
makes it necessary to use the explicit form of $\overline{\psi^{\theta}}$ 
as presented in section \ref{zwei}.

There is a growing literature of works on Landau-Zener
tunnelling in the presence of a mean-field nonlinearity and
its impact on the Bloch oscillation problem, see, e.g.,
\cite{BEC_asym,niuETAL}. 
In such works a similar systematical stabilisation (for $g<0$) or
destabilisation (for $g>0$) was predicted 
(and also measured, see \cite{BEC_asym}) for a {\em single} Landau-Zener
tunnelling event with various approximative models. This corresponds
to our results on the decay rates which describe directly the 
initial decay of the condensate via tunnelling, i.e., the
behaviour of $P_{\rm sur}(t)$ at short times 
that are not much larger than one Bloch period. 
At and close to the resonant tunnelling
peaks the problem is, however, more subtle because of
the strong interaction of Wannier-Stark levels [see Fig. \ref{fig:1}(a)], and
such a case was not treated in \cite{BEC_asym,niuETAL}.

\section{Conclusion} 
\label{concl} 

To summarise, we presented a method to numerically compute precise
decay rates for tunnelling problems within the framework of the
Gross-Pitaevskii equation. We adapted and improved the technique 
developed by one us in \cite{SP2004} for the more complicated
scenario of resonant tunnelling in the Wannier-Stark problem.
We showed that the mean-field nonlinearity leads to experimentally observable 
modifications in the tunnelling of resonance states from the 
periodical potential wells, even
in a regime where the kinetic and the periodic potential terms still
dominate the dynamics. The broadening and the shift of the resonant
tunnelling peaks define clear signatures for nonlinearity induced
effects. 

Our method can be extended -- with further system-specific improvements
in the propagation algorithm -- to treat even more complicated problems 
appearing in experiments with Bose condensates, e.g., the transport
of coherent matter within atomic wave guides. 
Finally, we can readily extend the proposed method to 
three spatial dimensions, with the only drawback of much larger 
numerical effort.

\ack 
S.W. is very grateful to Ennio Arimondo and Oliver Morsch for
valuable discussions on the experimental feasibility of the nonlinear 
Wannier-Stark system, and acknowledges support from the Alexander von 
Humboldt Foundation (Feodor-Lynen Program).

\end{document}